\documentclass[journal]{IEEEtran}
\usepackage{amsmath}
\usepackage{amssymb}
\usepackage{amsfonts}
\usepackage{graphicx}
\usepackage{subfigure}
\usepackage{algorithm}
\usepackage{algorithmic}
\usepackage{cite}
\usepackage{url}
\usepackage{multirow}
\usepackage{balance}
\usepackage{placeins}
\usepackage{float}
\newfloat{algorithm}{t}{lop}

\usepackage{mathtools}

\newcommand*{\bH}{\boldsymbol{H}}
\newcommand*{\bI}{\boldsymbol{I}}

\newcommand*{\bS}{\boldsymbol{S}}
\newcommand*{\bzero}{\boldsymbol{0}}

\newcommand*{\ba}{\boldsymbol{a}}

\newcommand*{\beff}{\boldsymbol{f}}

\newcommand*{\bh}{\boldsymbol{h}}

\newcommand*{\bv}{\boldsymbol{v}}
\newcommand*{\bw}{\boldsymbol{w}} 

\newcommand*{\by}{\boldsymbol{y}} 

\newcommand*{\bone}{\boldsymbol{1}}

\newcommand*{\rma}{\mathrm{a}}
\newcommand*{\rmb}{\mathrm{b}}

\newcommand*{\rmn}{\mathrm{n}}

\newcommand*{\rmr}{\mathrm{r}}

\newcommand*{\rmt}{\mathrm{t}}

\newcommand*{\rmx}{\mathrm{x}}
\newcommand*{\rmy}{\mathrm{y}}
\newcommand*{\rmB}{\mathrm{B}}
\newcommand*{\rmT}{\mathrm{T}}

\newcommand*{\rmpl}{\mathrm{pl}}
\newcommand*{\rminit}{\mathrm{init}}

\newcommand*{\rmA}{\mathrm{A}}
\newcommand*{\rmD}{\mathrm{D}}

\newcommand*{\jj}{\mathrm{j}}

\newcommand*{\SNR}{{\mathsf{SNR}}}

\newcommand*{\bbE}{\mathbb{E}} 
\newcommand*{\bbP}{\mathbb{P}} 

\newcommand*{\mcB}{\mathcal{B}}

\newcommand*{\mcO}{\mathcal{O}}

\newcommand*{\mcS}{\mathcal{S}}
\newcommand*{\mcT}{\mathcal{T}}

\newtheorem{proposition}{Proposition}

\newcommand*{\figwidthscale}{1}

\begin{document}
%
\title{Inverse Multipath Fingerprinting for Millimeter Wave V2I Beam Alignment}
%
%
%

\author{ 
Vutha Va, Junil Choi, Takayuki Shimizu, Gaurav Bansal, and~Robert W. Heath, Jr.
\thanks{Vutha Va and Robert W. Heath, Jr. are with the Wireless Networking and Communications Group, The University of Texas at Austin, TX 78712-1687 USA (e-mail: vutha.va@utexas.edu, rheath@utexas.edu).}
\thanks{Junil Choi is with the Department of Electrical Engineering, Pohang University of Science and Technology (POSTECH), Pohang, Gyeongbuk, Korea 37673 (e-mail: junil@postech.ac.kr).}
\thanks{Takayuki Shimizu and Gaurav Bansal are with TOYOTA InfoTechnology Center, U.S.A., Inc., Mountain View, CA 94043 USA (e-mail: tshimizu@us.toyota-itc.com, gbansal@us.toyota-itc.com).}
\thanks{This work is supported in part by the U.S. Department of Transportation through the Data-Supported Transportation Operations and Planning (D-STOP) Tier 1 University Transportation Center and by the Texas Department of Transportation under Project 0-6877 entitled ``Communications and Radar-Supported Transportation Operations and Planning (CAR-STOP)" and by a gift from TOYOTA InfoTechnology Center, U.S.A., Inc.}
}

\maketitle

\begin{abstract}
Efficient beam alignment is a crucial component in millimeter wave systems with analog beamforming, especially in fast-changing vehicular settings. 
This paper proposes a position-aided approach where the vehicle's position (e.g., available via GPS) is used to query the multipath fingerprint database, which provides prior knowledge of potential pointing directions for reliable beam alignment.
The approach is the inverse of fingerprinting localization, where the measured multipath signature is compared to the fingerprint database to retrieve the most likely position.
The power loss probability is introduced as a metric to quantify misalignment accuracy and is used for optimizing candidate beam selection. Two candidate beam selection methods are developed, where one is a heuristic while the other minimizes the misalignment probability. 
The proposed beam alignment is evaluated using realistic channels generated from a commercial ray-tracing simulator. 
Using the generated channels, an extensive investigation is provided, which includes the required measurement sample size to build an effective fingerprint, the impact of measurement noise, the sensitivity to changes in traffic density, and beam alignment overhead comparison with IEEE 802.11ad as the baseline. Using the concept of beam coherence time, which is the duration between two consecutive beam alignments, and parameters of IEEE 802.11ad, the overhead is compared in the mobility context. The results show that while the proposed approach provides increasing rates with larger antenna arrays, IEEE 802.11ad has decreasing rates due to the larger beam training overhead that eats up a large portion of the beam coherence time, which becomes shorter with increasing mobility.
\end{abstract}


%
\IEEEpeerreviewmaketitle

\section{Introduction}
%
%
%
%

Communication enhances advanced driver assistance systems, enables a wide range of infotainment options, and paves the way towards fully automated driving. Vehicular automation capabilities rely heavily on perception sensors such as camera, radar, and LIDAR \cite{Levinson2011}. Sharing the data from these sensors between vehicles, or with the infrastructure, can extend the sensing range as well as provide redundancy in case of sensor failures. High data rate links with the infrastructure also enable map updates, edge-network control of vehicles, and a wider range of infotainment services. 
Unfortunately, the data rate requirements are increasing beyond what can be provided by the fourth generation (4G) cellular or Dedicated Short-Range Communication (DSRC) solutions \cite{Kenney2011,Aranti2013,Junil2016}.

Millimeter wave (mmWave) is a means to provide high data rates in vehicle-to-everything (V2X) settings \cite{mmwave-vehicular-survey,Junil2016}, thanks to the large spectral channels \cite{mmwave-vehicular-survey}. The need for adaptive antennas to overcome the shrinking antenna aperture at high frequencies is one of the main challenges in using mmWave \cite{mmwave-vehicular-survey,Junil2016}. Due to the use of sharp beams and the high susceptibility to blockage, beam pointing directions have to be aligned and readjusted according to the changes in the environment. 
The high mobility in the vehicular context will require frequent beam realignment. Therefore, fast and efficient beam alignment is crucial in enabling high data rate mmWave vehicular communications. 

Prior knowledge of the propagation environment can help reduce the beam alignment overhead. This paper focuses on the use of multipath fingerprints, which are the long-term multipath channel characteristics associated with locations. The term ``fingerprint" originates from the localization literature \cite{Bahl2000,Lakmali2008,Kupershtein2013}, 
where the main premise is that channel characteristics are highly correlated with locations. 
In fingerprinting-based localization methods, there is a fingerprint database, which records fingerprints at different locations in the area of interest. When a terminal wants to localize itself, it first performs radio frequency (RF) channel measurements to obtain the multipath fingerprint at the current location. The obtained fingerprint is then matched against the fingerprints in the database and the output location is computed based on the matched fingerprints in the database that are ``closest" to the observed fingerprint. 
Considering the availability of position information in the vehicular context, we propose to use this idea in inverse.
Localization is an important task in driving automation, where
vehicles position themselves via a suite of sensors including Global Positioning System (GPS), cameras, and LIDAR \cite{Ward2016}. 
This position information can be used to query the fingerprint database which is indexed by location to determine beam directions that are likely to provide satisfactory link quality. 
Because of the sparsity of mmWave channels \cite{Rappaport2013}, 
potential directions are expected to be highly concentrated, and thus we expect a large reduction in the beam training overhead. 

The objective of this paper is to develop an efficient beam alignment method suitable for a vehicle-to-infrastructure (V2I) setting. Our contributions are summarized as follows.
\begin{itemize}
\item We propose a novel and efficient beam alignment method using multipath fingerprints. This paper focuses on an offline learning setting where there is a dedicated period of time for building the database before it is used for efficient beam alignment. We provide a numerical example to show its extensibility to an online learning setting.
\item Two types of fingerprints (Types A and B) are proposed, which differ in how measurements are collected and stored. Type A assumes each contributing vehicle performs an exhaustive search over all beam pairs (so that correlation between beam pairs can be captured), while Type B collects only a fraction of the exhaustive search at a time. Type A stores the raw received power, while Type B only stores the average received power of each beam pair. This provides flexibility for actual implementations, where data collection and storage cost must be met.
\item We introduce the power loss probability as a metric for evaluating the beam alignment accuracy. This metric leads to a mathematical framework for optimizing the candidate beam pair selection. 
We propose two beam pair selection methods, where one is a heuristic and the other is a solution that minimizes the misalignment probability. 
\item We provide an extensive numerical investigation, which includes the training sample size to build the fingerprint database, beam training overhead comparison with IEEE 802.11ad, and the sensitivity to changes in vehicular traffic density. For the overhead comparison, we leverage the concept of beam coherence time \cite{Va:Impact-of-beamwidth:2016} to quantify the beam training cost in the vehicular context. We use realistic channels generated from a commercial ray-tracing simulator, Wireless InSite \cite{Remcom_WI}, in all our results. 
\end{itemize}
We note that while we emphasize the V2I context in this work, the approach can also be applied to general cellular settings. An additional challenge is in how to determine the orientation of the antenna array of the user equipment (which is needed to translate angles of arrival and departure (AoAs/AoDs) to beam indices). This is not as important for vehicles because the array is fixed on the vehicle (e.g., the roof) and the orientation can be determined from the heading of the vehicle. 

Beam alignment is a subject of intense research because of its importance in enabling mmWave communications. Here, we summarize and compare with relevant work in the context of analog beamforming, where both the transmitter and receiver have only one RF chain. We group existing solutions into four categories: beam sweeping \cite{Wang:Beam-codebook-based-beamforming-protocol:09,Hur:Millimeter-wave-beamforming-for-wireless-backhaul:13,Hosoya:MIDC-a-novel-beamforming-technique:14}, AoAs/AoDs estimation \cite{Kim2015a,Duan2015,Marzi2016}, blackbox optimization \cite{Li2013,Li2014b,Kadur2016}, and side information \cite{Prelcic2016,Nitsche2015,Ali2017,Kim:Enabling-Gigabit-services-for-IEEE802-11ad:13,Va:beam-design:2016,Abbas2016,Capone2015,Aviles2016}. 

Beam sweeping involves a set of beam measurements designed to cover all possible pointing directions, whose simplest form is the exhaustive search. 
It is simple and robust because it makes few assumptions on the channel. It only requires that the spatial channel does not change during the sweeping time. This approach has been adopted in existing mmWave standards such as IEEE 802.15.3c \cite{15.3c} and IEEE 802.11ad\cite{802.11ad} for indoor use cases. The brute force search measures all transmit and receive beam pair combinations resulting in quadratic complexity in the beam codebook size. Our approach reduces the search time by prioritizing the most promising directions identified from the fingerprints. The beam search can also be done hierarchically \cite{Wang:Beam-codebook-based-beamforming-protocol:09,Hur:Millimeter-wave-beamforming-for-wireless-backhaul:13,Hosoya:MIDC-a-novel-beamforming-technique:14}, but the search time does not change much because large spreading is needed to use wide beams in the initial stage \cite{Liu2017}. 


AoA/AoD estimation leverages the sparsity of mmWave channels to reduce the number of measurements required compared to beam sweeping. For example, compressive sensing is used in \cite{Kim2015a,Duan2015} while an approximate maximum likelihood estimator is derived using the channel structure directly in \cite{Marzi2016}. Compressive measurements have to overcome the lack of antenna gain during the measurement phase, unlike our approach that uses narrow beams for the beam training. 

Blackbox optimization is another approach to efficiently explore the beam directions \cite{Li2013,Li2014b,Kadur2016}. This framework is based on the premise that there is some structure (e.g., smoothness) of the objective function (i.e., the received power here), and thus one does not have to blindly search all the beam pair combinations. 
This approach uses narrow beams in the initial search like our method, but it requires a larger number of feedbacks to navigate the search region and this feedback overhead will be the bottleneck in reducing the search time. 

The final category uses side information available from sensors (including communication systems at other frequencies).  Radar information is used in \cite{Prelcic2016}, and information from lower frequencies is used in \cite{Nitsche2015,Ali2017}. More related to our work are those that use position information \cite{Kim:Enabling-Gigabit-services-for-IEEE802-11ad:13,Va:beam-design:2016,Abbas2016,Capone2015,Aviles2016}. 
The work in \cite{Kim:Enabling-Gigabit-services-for-IEEE802-11ad:13,Va:beam-design:2016,Abbas2016} uses position to determine beam directions. This can eliminate the beam training overhead but it only works when the LOS path is available. More elaborate channel models with LOS obstruction have been investigated in \cite{Capone2015,Aviles2016}. It is proposed in \cite{Capone2015} to memorize successful beam configurations observed in the past but no detail is given on how to rank those configurations in terms of their likelihood to provide a good link. Omnidirectional antennas at the users are assumed in \cite{Capone2015}, which may be impractical for mmWave communications. In \cite{Aviles2016}, a heuristic is proposed for a hierarchical beam search with the help of position and multipath database. Only the search in the azimuth was considered and horn antennas were assumed.  
Our proposed approach is in this category, where we use position information and multipath fingerprints. Unlike \cite{Kim:Enabling-Gigabit-services-for-IEEE802-11ad:13,Va:beam-design:2016,Abbas2016}, by leveraging multipath fingerprints, our approach can work in both LOS and non-LOS (NLOS) channels. Different from \cite{Capone2015,Aviles2016}, the proposed beam training uses narrow beams at both the roadside unit (RSU) and the vehicle. Also, we provide a mathematical framework to rank beam pointing directions from past measurements in the fingerprint database and both azimuth and elevation are considered. 


The rest of the paper is as follows. The system model is described in Section~\ref{sec:system_model}. In Section~\ref{sec:inverse_fingerprint_beam_alignment}, we define the multipath fingerprints and explain how the proposed beam alignment works. 
We provide an analytical framework for quantifying beam alignment accuracy in Section~\ref{sec:analysis}.
In Section~\ref{sec:selection_method}, we present our beam candidate selection methods using the fingerprints. 
An extensive numerical investigation is provided in Section~\ref{sec:numerical_result}. Finally, the paper is concluded in Section~\ref{sec:conclusion}.

\section{System Model} \label{sec:system_model}
This paper proposes an efficient beam alignment method for V2I communications in an urban street canyon environment with high traffic density, where LOS is often unavailable and represents a challenging scenario for beam alignment. 
While the methodology of the beam training is general, we test the approach using data from
a commercial ray-tracing simulator, Wireless InSite \cite{Remcom_WI}. 
The evaluations of the proposed method are conducted via post-processing in MATLAB. 

\subsection{Channel Model} \label{sec:channel_model}
\begin{figure}
\centering
\includegraphics[width=0.9 \columnwidth]{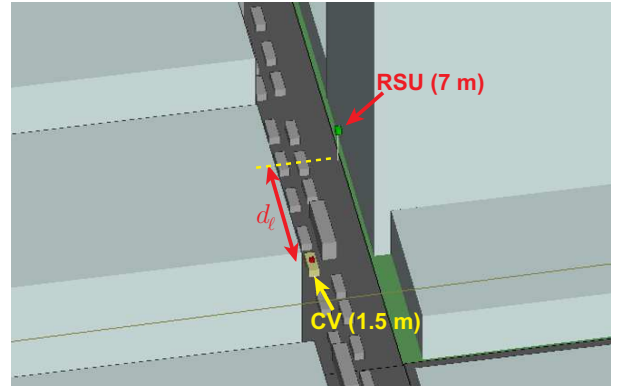}
\caption{Ray-tracing simulation environment. The numbers in the parenthesis are the antenna heights.}
\label{fig:WI_sim_scenario}
\end{figure}

The simulation environment is shown in Fig. \ref{fig:WI_sim_scenario}, which is an urban street with two lanes. All the buildings are made of concrete (relative permittivity $\epsilon_\rmr=5.31$ and conductivity $\sigma=0.8967$ S/m \cite[Table 3]{ITU-building-material}), and the road surfaces are made of asphalt ($\epsilon_\rmr=3.18$ and $\sigma=0.3338$ S/m \cite{Li1999}). The surface root mean square roughness is set to 0.2 mm for concrete and 0.34 mm for asphalt \cite[Table 1]{Li1999}. We allow up to two reflections and one diffraction. 
We simulate two types of vehicles represented by metal boxes (made of perfect electric conductor which is predefined in Wireless InSite \cite{Remcom_WI}): cars ($1.8\,\mathrm{m}\times 5 \,\mathrm{m} \times 1.5 \,\mathrm{m}$) and trucks ($2.5\,\mathrm{m}\times 12 \,\mathrm{m} \times 3.8 \,\mathrm{m}$). 
The cars-to-trucks ratio is 3:2. 
The RSU is placed on the right side, and a car on the left lane is selected as the communicating vehicle (CV). 
The antenna heights are 7 m and 1.5 m for the RSU and the CV (on its roof), respectively.
Because trucks are taller, they could block the LOS path between the CV and the RSU. The carrier frequency is set to 60 GHz.

To imitate the dynamic blockage environment, we simulate multiple snapshots of the scenario where vehicles are independently and randomly placed in each snapshot. The gap between vehicles (i.e., from the front bumper to the rear bumper of the heading vehicle) $\zeta$ is randomly determined from the Erlang distribution \cite{Al-Ghamdi2001} given by
\begin{align}
f_\zeta (\zeta) = \frac{(\kappa\mu_\zeta)^\kappa}{(\kappa-1)!}\zeta^{\kappa-1}e^{-\kappa\mu_\zeta \zeta},
\label{eq:pdf_veh_gap}
\end{align}
where $\kappa$ is the shape parameter and $1/\mu_\zeta$ is the mean gap. Following \cite{Mase2008}, $\kappa=6$ and $\mu_\zeta=0.209$ are used in our simulation, which produces an average gap of $4.78$ m. 
Since multipath fingerprints are associated with locations, we need to generate multiple channels at a given location. To do this, the CV is placed at a longitudinal distance $d_\ell$ from the RSU (see Fig. \ref{fig:WI_sim_scenario}), where $d_\ell$ is uniformly drawn from $[d_0-\sigma_d,d_0+\sigma_d]$ for some mean distance $d_0$ and some grid size $2\sigma_d$. $d_0=30$ m and $\sigma_d=2.5$ m are used when generating the channels. When applying our method, all points within this range are treated as being in the ``same" location bin indexed by $d_0$. By discretizing the location this way, 
we can provide some resilience to errors in position information estimated by the vehicle. The edge effect can be mitigated by defining overlapping location bins.

We note that no mobility is considered during the beam training. The beam training duration for the proposed method is sub-millisecond (see Section \ref{sec:numerical_comparison}), so the displacement of vehicles during the beam training is negligible. For example, when $16\times 16$ arrays are used the training duration is about 150 $\mu$s, so the displacement is only $3$ mm even with a speed of 20~m/s (high in urban scenarios). Our proposed beam alignment can be considered as an initial link establishment, and after which a beam tracking method such as \cite{Gao2017} could be used to maintain the link to further reduce the overhead. Each channel sample corresponds to a random and independent placement of vehicles as described earlier in a per snapshot basis. These different snapshots produce blocking dynamics, where a good path in a snapshot could be useless due to blockage in another snapshot even if in both snapshots the CV is at the same location bin. 

The output from the ray-tracing simulation is combined with the geometric channel model to obtain the corresponding channel matrix. A ray-tracing simulation outputs a number of rays, each corresponding to a distinct propagation path. The information associated with each ray includes the received power, the delay of the path, the phase, the AoA, and the AoD. Denoting $(\cdot)^*$ the conjugate transpose, $N_\rmt$ and $N_\rmr$ the numbers of transmit and receive antennas, $L_\mathrm{p}$ the number of rays, $\alpha_\ell$ the complex channel gain, $\tau_\ell$ the delay, $\theta_\ell^{\rmA}$ and $\theta_\ell^{\rmD}$ the 
elevation AoA and AoD, $\varphi_\ell^{\rmA}$ and $\varphi_\ell^{\rmD}$ the azimuth AoA and AoD of the $\ell$-th ray, $g(\cdot)$ the combined effect of lowpass filtering and pulse shaping, $B$ the system bandwidth, $T=1/B$ the symbol period, and $\ba_\rmr(\cdot)$ and $\ba_\rmt(\cdot)$ the receive and transmit steering vectors, the channel can be written as
\begin{align}
\label{eq:channel_model}
\bH[n]\! =\! \sqrt{\! N_\rmr N_\rmt}\! \sum_{\ell=0}^{L_\mathrm{p}-1}\!\!  \alpha_\ell g(nT-\tau_\ell) \ba_\rmr(\theta_\ell^{\rmA},\varphi_\ell^{\rmA})
\ba_\rmt^*(\theta_\ell^{\rmD},\varphi_\ell^{\rmD}). 
\end{align}
The raised cosine filter with a roll-off factor of 0.1 is assumed for the pulse shaping filter. The total number of rays is $L_\mathrm{p}=25$. We use the Full3D model with Shooting and Bouncing Ray (SBR) tracing mode in Wireless InSite. The ray spacing is $0.25^\circ$, which means the simulator shoots hundreds of rays and determines which of them form valid propagation paths and records the strongest 25 of them. The power gap among these 25 rays is more than 20 dB, and thus there is little value in keeping more rays. We also note that since we use ray-tracing, there is no need to specify the number of clusters or rays per cluster as typical of stochastic channel models. The ray-tracing will determine all relevant propagation paths. In fact, some of the 25 rays can be thought of as belonging to the same cluster in the sense that they have similar delays and AoAs. Note that by using ray-tracing, we ensure that the channels are spatially consistent, which is a feature not available in most stochastic channel models. 

We use uniform planar arrays (UPAs) at both the transmitter and the receiver. Let $\Omega_\rmy = k d_\rmy\sin(\theta)\sin(\varphi)$, $\Omega_\rmx = k d_\rmx\sin(\theta)\cos(\varphi)$, $k=2\pi/\lambda$ be the wave number, $\otimes$ denote the Kronecker product, $N_\rmx$ and $N_\rmy$ be the numbers of elements along the x- and y-axis, and $d_\rmx$ and $d_\rmy$ be the element spacing in the x- and y-direction, the steering vector is
\begin{align*}
\ba(\theta,\varphi) = \frac{1}{\sqrt{N_\rmx N_\rmy}}
\begin{bmatrix}
1 \\
e^{\jj \Omega_\rmy}  \\
\vdots \\
e^{\jj (N_\rmy-1)\Omega_\rmy} 
\end{bmatrix} \otimes \begin{bmatrix}
1 \\
e^{\jj \Omega_\rmx} \\
\vdots \\
e^{\jj (N_\rmx-1)\Omega_\rmx}
\end{bmatrix}.
\end{align*}
In this paper, we use $d_\rmx=d_\rmy=\lambda/2$. 

\subsection{Received Signal Model}
We assume a time division duplex (TDD) system with analog beamforming, which uses only one RF chain. 
We assume the symbol timing is synchronized to the first path (shortest delay). This means that paths with larger delay are not likely synchronized to the sampling timing and the energy will leak to adjacent symbols, with the leak amount determined by the combined filter response $g(\cdot)$. The received signal when the $i$-th beam pair is used can be written as the time-domain convolution between the transmit signal and the effective channel seen through the $i$-th beam pair $h_i[n]$, i.e., 
\begin{align}
& y_i[k] = \nonumber \\&
\sqrt{P_\rmt} \sum_{n=0}^{L-1} s[k-n] \underbrace{\sum_{\ell=0}^{L_\mathrm{p}-1} g(nT+\tau_0-\tau_\ell)\bw_{r(i)}^* \bH_\ell \beff_{t(i)}}_{h_i[n]} + v_i[k],
\label{eq:Rx_signal_model}
\end{align}
where $\bH_\ell = \sqrt{N_\rmr N_\rmt} \alpha_\ell e^{\jj \beta_\ell} \ba_\rmr(\theta_\ell^{\rmA},\varphi_\ell^{\rmA}) \ba_\rmt^*(\theta_\ell^{\rmD},\varphi_\ell^{\rmD})$, $L$ is the channel length, $P_\rmt$ is the transmit power, $s[k]$ is the known training signal,
$r(i)$ and $t(i)$ denote the mapping of the beam pair index $i$ to the combiner and beamformer vector indices, and $v_i[k]$ is the zero mean complex Gaussian noise $\mathcal{CN}(0,\sigma_v^2)$. 
Let $\by_i=\begin{bmatrix}
y_i[0] & y_i[1] & \dots & y_i[L-1]
\end{bmatrix}^\rmT$, $\bh_i=\begin{bmatrix} h_i[0] & h_i[1] & \dots & h_i[L-1] \end{bmatrix}^\rmT$ and $\bv_i=\begin{bmatrix}
v_i[0] & v_i[1] & \dots & v_i[L-1]
\end{bmatrix}^\rmT$, we can rewrite the received signal in a matrix form as
\begin{align}
\by_i = \bS \bh_i + \bv_i, 
\end{align}
where $\bS$ is the $K\times L$ circularly shifted training sequence with $K\ge L$ the training sequence length. The channel can be estimated using a least-square approach as
\begin{align}
\hat{\bh}_i & = \bS^\dagger \by_i
\\ & = \bh_i + \bS^\dagger \bv_i
\end{align}
where $\bS^\dagger = (\bS^*\bS)^{-1}\bS^*$ is the pseudo-inverse. When using a training sequence with good autocorrelation properties like Zadoff-Chu or Golay sequences, $\bS^*\bS=K\bI$ and the estimation error $\tilde{\bv}_i=\bS^\dagger\bv_i$ can be modeled as $\mathcal{CN}\left( \bzero, \frac{\sigma_v^2}{K}\bI  \right)$ \cite{wirelessLabTextbook}. We refer to \cite[Chapter 5]{wirelessLabTextbook} for more details. 
Note that the received powers in the fingerprints have to be measured using the same $P_\rmt$ (or scaled appropriately if different $P_\rmt$ are used). 
In our simulations, we use $K=512$ which is motivated by the training sequence used in IEEE 802.11ad. The Channel Estimation Field (CEF) of IEEE 802.11ad frame consists of two training sequences of length 512 which can be used to perform two independent channel estimations \cite{802.11ad}. The actual channel length $L$ varies for different snapshots of the ray-tracing simulation and can be larger than 512. Since the powers of those paths with large delays are observed to be negligible compared to paths with short delays, we truncate the channel to get $L=512$.


The vector $\bw_{r(i)}$ and $\beff_{t(i)}$ are selected from the receiver codebook $\mathcal{W}$ and transmitter codebook $\mathcal{F}$, respectively.
We assume UPAs are used at both the CV and the RSU.  
The beams in the codebook are generated using progressive phase shift \cite{Orfanidis2014} between antenna elements. 
The main lobe directions of the generated beams are separated by their 3dB-beamwidth. This ensures that the array gain fluctuates less than 3 dB over the entire field of view of the antenna array. We note that our proposed approach does not depend on this specific codebook, and any other codebook designs can replace the one used here. 


\section{Inverse Fingerprint Beam Alignment} \label{sec:inverse_fingerprint_beam_alignment}
The main idea of the proposed approach is to leverage prior knowledge to identify promising beam directions and only train those directions.
The prior knowledge is 
obtained from past observations in the database, and some of these paths may not exist in the current channel due to blockage by a truck for example. Therefore, beam training among the beam directions identified from the database is still required. The beam training here has much lower overhead than conventional methods because a large number of unlikely directions have already been eliminated using the database. In this section, we will define fingerprints and explain how the database is constructed. Then, we will describe the proposed approach for beam alignment.  

\subsection{Multipath Fingerprint Database} \label{sec:multipath_fingerprint}
In general, a fingerprint refers to some characteristics of the channel at a given location. These characteristics could be the received signal strengths from different access points \cite{Lakmali2008} or the multipath signature of the channel from an access point \cite{Kupershtein2013}. In this work, a fingerprint refers to a set of 
received powers of different pairs of transmit and receive beams at a given location.
We note that the location here refers to a grid $[d_0-\sigma_d,d_0+\sigma_d]$ so that the system can tolerate position information error used to query the fingerprint database. 

We define two types of fingerprints, which differ by how measurement data are collected and stored. The first type, called Type A, requires that the contributing vehicle perform a full exhaustive beam measurements over all beam pairs. This ensures that the measurements over the different pairs happen within a channel coherence time so that the spatial channel does not change. This way, the fingerprint captures the correlation between the different beam pairs, i.e., whether they tend to have similar received power or not. For Type A fingerprints, the raw measurement samples are stored. One could store all the measurements of all the beam pairs from each contributing vehicle, but if memory storage is of concern, only the measurements of the top-$M$ beam pairs (ranked by the received power) can be stored. This makes sense because most of the beam pairs have negligible received power (since their main beam directions do not point along any propagation paths), and thus there is negligible information gain in keeping all beam pairs. In our simulation, we use $M=100$. The average power difference between the strongest beam pair and the 100th strongest over the channel samples used is 22.2 dB. 
An example of Type A fingerprints is shown in Table \ref{tab:raw_fingerprint_example}. In this example, there are $N$ observations collected by $N$ contributing vehicles. 

Type B fingerprints do not require that the measurement of all the beam pair combinations be completed within a channel coherence time. This less restrictive data collection reduces the burden on individual vehicles contributing to building the database, as they do not need to commit to conducting a full scan of the exhaustive search and could contribute as many beam measurements as their time allows. A simple method would be to do a round-robin over the beam pair combinations. For example, assume that each vehicle can do only $1/4$ of the full exhaustive search. The RSU then divides the set of all beam pairs into four disjoint sets and assign these sets sequentially to subsequent contributing vehicles to collect the measurement data.
The disadvantage is that now the correlation between the beam pairs cannot be easily captured. Only the average received powers (the average is computed in linear scale) of the beam pairs are stored. An example of Type B fingerprints is shown in Table \ref{tab:fingerprint_example}. We note that the sample averages can be computed recursively, and thus there is no need to temporally store the raw samples when collecting Type B fingerprints.

\begin{table}
\centering
\caption{An example of Type A fingerprints. 
For each contributing vehicles, the measurements of the top-$M$ beam pairs are stored. In each cell, the integer at the top is the beam pair index and the number at bottom is the received power. 
}
\label{tab:raw_fingerprint_example}
\begin{tabular}{|c|c|c|c|c|} \hline 
Observation No. & Best & 2nd best & $\dots$ & $M$-th best \\ \hline
\multirow{2}{*}{1} &  5  & 159 & $\dots$ & 346 \\
 & -64.5 dBm & -69.2 dBm & $\dots$ & -95.8 dBm \\ \hline
\multirow{2}{*}{2} &  159  & 263 & $\dots$ & 354 \\
  & -70.4 dBm & -72.6 dBm & $\dots$ & -97.1 dBm \\ \hline 
  ... & ... & ... & ... & ... \\ \hline
\multirow{2}{*}{$N$} &  5  & 258 & $\dots$ & 2 \\ 
   & -66.4 dBm & -68.1 dBm & $\dots$ & -82.6 dBm \\ \hline 
\end{tabular}
\end{table}

\begin{table}
\centering
\caption{An example of Type B fingerprints at a location bin. The average received power for each beam pair is recorded.  
}
\label{tab:fingerprint_example}
\begin{tabular}{|c||c|c|c|} 
\hline
Beam pair index & 1 & 2 & $\dots$ \\ \hline
Average received power & -92.3 dBm & -73.5 dBm & $\dots$  \\ \hline
\end{tabular}
\end{table}


We focus on an offline learning setting to build the database in this work. 
The idea can readily be implemented online as shown in Section \ref{sec:online_learning}. 
In the proposed approach, the database is built and maintained by the RSU. By offline learning, we mean there is a dedicated period of time that is used to collect data to build the fingerprint database before it is exploited for efficient beam alignment. During this period, each contributing vehicle conducts beam training with the RSU. By having the vehicle be the transmitter during the beam training, there is no feedback. The RSU can measure the received power for each of the beam pairs and collect these observations to build the fingerprint database.  
In the case that each contributing vehicle can perform a full exhaustive search over all beam pair combinations in the codebook, we obtain Type A fingerprints. 
If each contributing vehicle cannot perform a full exhaustive search, we get Type B fingerprints. 

We now discuss the cost for building and storing the database. The database can be collected in the initial stage of the system deployment. The RSU can request vehicles passing by its coverage to conduct beam training. 
Most modern vehicles are GPS equipped either for navigation or for safety message (which includes position, speed, and heading among other information) dissemination such as in DSRC. Thus, it is fair to assume that any vehicles equipped with mmWave communication also have positioning capability, and all mmWave communication capable vehicles can contribute to build the database. As shown in Section \ref{sec:req_sample_size}, around 250 full exhaustive beam measurements 
are enough to get a fully functioning database. 
In a dense urban traffic setting, 
this could be done within an hour if not less (e.g., the number of vehicles passing through an urban road segment was more than 100 per lane in 15 minutes as reported in the NGSIM Lankershim dataset \cite{NGSIMdata}). We note that if exhaustive search is used as the beam alignment method when the database is not available \cite{Liu2017}, the data collection is essentially free since the vehicles will need to conduct the exhaustive search to establish the link during this stage. 

The next question is how frequent the database needs to be updated. The main premise of the proposed method is to learn the long-term multipath information from the database, which is the propagation directions that depend on the geometry of the environment such as the road structure and buildings. The change in the traffic density can affect the relative importance of different paths, but as shown in Section \ref{sec:sensitivy_to_traffic_density}, if the database is collected in a dense traffic, it will also work well in light traffic conditions. Thus, we expect the database collected in high traffic density to be valid for a long period of time such as weeks or even months if there is no construction in the surrounding buildings and road structure. Of course, once a functioning database has been established, it can be reinforced by requesting idle vehicles passing by to conduct beam measurements and replace the older data with the newly collected ones. This update can be done at a slow pace since we expect the database to change slowly. 

Finally, we consider the storage requirement. For Type A fingerprints, using $4$ bytes for one received power and $2$ bytes for one beam index, the total storage of Table \ref{tab:raw_fingerprint_example} is $6NM$ bytes. Assuming $N=250$, $M=100$ and 200 location bins per RSU coverage (1 m bin size and 200 m RSU coverage), it requires about 30 MB. Type B requires even less data storage. This amount of data can be easily stored in any modern devices. Therefore, we conclude that storage is not at all a problem.

\subsection{Proposed Beam Alignment} \label{sec:proposed_beam_alignment}
\begin{figure}
\centering
\includegraphics[width=0.9\columnwidth]{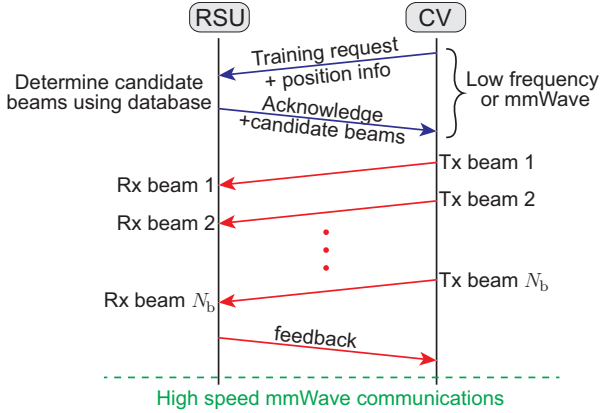}
\caption{Timing diagram of the proposed inverse fingerprint beam alignment method. The CV sends a beam training request with its position to the RSU. The RSU acknowledges and provides a list of candidate beam pairs for training by querying the fingerprint database using the CV's position. Beam training can then start following the list. At the end, the RSU feeds back the best observed beam to the CV and high speed mmWave communications can begin.}
\label{fig:inverse_fingerprint_beam_alignment}
\end{figure}

Fig. \ref{fig:inverse_fingerprint_beam_alignment} shows how the proposed beam alignment works. The procedure is initiated by a CV by transmitting a beam training request using a low-frequency system such as DSRC or mmWave communication with a large spreading factor. Within this request packet, the CV also includes its current position, which is available from some localization sensors equipped on the vehicle (e.g., GPS or other more advanced positioning method based on LIDAR and 3D map \cite{Ward2016}). Upon receiving the request, the RSU will use the position information to query its database to obtain the fingerprint associated with that location. Using the fingerprint, the RSU determines the candidate beam pairs that are likely to provide a satisfactory link connection. The RSU then responds back with an acknowledgment to allow the beam training and also provides a list of candidate beam pairs. 
Beam pair selection methods will be described in Section \ref{sec:selection_method}. 
The CV can now proceed to perform the beam training following the list provided by the RSU. Note that the list consists of beam pairs, and there is no need to do exhaustive measurements over all combinations of transmit and receive beams in the list. 
Upon completing the list, the RSU provides a feedback indicating the best beam index to the CV. This feedback ends the beam alignment and high data rate mmWave communication can start. 

Several remarks on the proposed method are now provided.

{\it Remark 1:} The position information here does not have to be highly accurate. It only needs to be accurate enough to identify the location bin index of the fingerprints. In our simulation, this bin size is 5 m. Edge effects can be avoided by having overlapping location bins. 

{\it Remark 2:} The proposed method allows graceful degradation as the number of beam pairs trained $N_\rmb$ decreases because the alignment accuracy decreases probabilistically with $N_\rmb$ and there is no hard threshold on $N_\rmb$. See Fig. \ref{fig:rate_plot} for an example of how the average rate changes with $N_\rmb$.
This is a desirable feature that allows the tradeoff between latency in link establishment and accuracy of the beam alignment. 

{\it Remark 3:} Our method performs the beam training using only narrow beams, which provides several advantages. Narrow beams provide high antenna gain and are more resilient to Doppler spread \cite{Va:Impact-of-beamwidth:2016}. 
Also, methods employing wide quasi-omni beams can suffer from antenna gain fluctuation because it is challenging to produce wide beamwidths with low gain fluctuation \cite{Hosoya:MIDC-a-novel-beamforming-technique:14}. 

{\it Remark 4:} By having the CV transmit and the RSU receive during the beam training, the RSU obtains beam measurements for free, i.e., without any feedback from the CV. These measurements can be used for updating the database. 

\section{Quantifying Beam Alignment Accuracy} \label{sec:analysis}

This section defines a metric for measuring the beam alignment accuracy, which allows us to compare different candidate beam pair selection methods and can serve as a framework for optimizing the method to select candidate beam pairs for training in Section \ref{sec:selection_method}. 
The definition here is defined assuming the measurement noise is negligible. In the presence of measurement noise, the metrics computed using \eqref{eq:def_power_loss} and \eqref{eq:power_loss_def1} will be less accurate and can affect the beam pair selection and degrade the beam alignment accuracy. 
We investigate the effect of noise numerically in Section \ref{sec:numerical_result_noise_impact} and show that the degradation due to noise is negligible if the transmit power is not too low. 


The power loss is defined as the ratio between the received power of the optimal beam pair and the pair selected by the beam alignment method indexed by $s$. Let $\mcB$ be the set of all beam pair combinations, and $\gamma_\ell=\|\bh_\ell\|^2$ be the received power of the $\ell$-th beam pair. The power loss can be written as 
\begin{align}
\xi = \frac{\max_{k\in\mcB}\gamma_{k}}{\gamma_{s}}.
\label{eq:def_power_loss}
\end{align}
If noise is negligible, the strongest beam pair will be selected after the beam training so that $\gamma_s=\max_{i\in\mcS}\gamma_{i}$, where $\mcS\subset\mcB$ is the set of candidate beams selected for beam training. 
The power loss probability is defined as the probability that $\xi>c$ for some $c\ge1$, i.e., 
\begin{align}
P_\rmpl(c,\mcS) & = \bbP[\xi> c] 
\label{eq:power_loss_def1}
\\ &= \bbP\left[ \max_{k\in\mcB}\gamma_{k}> c  \max_{i\in\mcS}\gamma_{i} \right].
\label{eq:power_loss_def2}
\end{align}
We note that since $\mcS$ is defined as a subset of $\mcB$, the definition in \eqref{eq:power_loss_def2} is always well-defined.

\section{Candidate Beam Pair Selection Methods} \label{sec:selection_method}
This section proposes two methods to select candidate beam pairs for beam training using the information in the fingerprint database. The objective of the selection methods is to maximize the received power of the finally selected beam pair for a given beam training budget of $N_\rmb$. The first approach is a heuristic, while the second one minimizes the misalignment probability defined in Section \ref{sec:analysis}. 
The heuristic is intended to be used with Type B fingerprints, and the other method is to be used with Type A fingerprints.

\subsection{Selection by Ranking Average Received Powers}
This method is based on the simple intuition that we should choose candidate beam pairs with the highest expected received power. 
The proposed approach is to first rank the beam pairs by their average received powers in descending order and select the highest $N_\rmb$ pairs for beam training.
Note that this metric can balance the selection of opportunistic paths because such paths will occasionally have high received power so that their average received powers are relatively high. If the occurrence of the opportunistic path is high enough, its average received power will be larger than a path that always has moderate received power and the opportunistic path is selected. If the occurrence is rare, the average received power of the opportunistic path is low and the path with always moderate received power is selected by this method. Thus, this metric can balance the risk and gain to some extent. We call this method AvgPow.

\subsection{Selection by Minimizing the Misalignment Probability} \label{sec:beam_sel_minMisProb}
Since the objective of beam alignment is to maximize the received power, an indirect way to achieve that is to choose the beam pair to minimize the misalignment probability, which is the power loss probability $P_\rmpl(c=1,\mcS)$. 
For a given training budget of $N_\rmb$, the problem can be formulated as a subset selection problem given by
\begin{align}
\underset{\mcS\subset \mcB}{\text{minimize }} & \,\, P_\rmpl(1,\mcS) \label{eq:powLoss_min_prob} \\ 
\text{subject to } & \,\, |\mcS|=N_\rmb . \nonumber
\end{align}
Here, $|\mcS|$ denotes the cardinality of the set $\mcS$.
This is a subset selection problem, which is combinatoric in nature and is difficult to solve in general, especially when $|\mcB|$ is large. 
Fortunately, the structure of $P_\rmpl(1,\mcS)$ allows an efficient solution. Note that the problem \eqref{eq:powLoss_min_prob} is equivalent to maximizing $\bar{P}_\rmpl(1,\mcS)=1-P_\rmpl(1,\mcS)$. Since $\bar{P}_\rmpl(1,\mcS)$ is a modular function, the greedy solution given in Algorithm \ref{alg:min_power_loss_algorithm} is optimal \cite[Theorem 7]{Edmonds1971}. This is a well-known result that has been reported in different forms in the literature (see \cite{Calinescu2011} and references therein.)


\begin{proposition}
\label{prop:modularity}
$\bar{P}_\rmpl(1,\mcS)$ is modular.
\end{proposition}
\begin{IEEEproof}
Using the definition of power loss probability in \eqref{eq:power_loss_def2} with $c=1$, we have
\begin{align}
&P_\rmpl (1,\mcS)  = \bbP\left[ \max_{k\in\mcB} \gamma_k > \max_{i\in\mcS}\gamma_i \right]
\label{eq:additivity_proof1}
\\ & = \sum_{\ell\in\mcB} \bbP\left[ \gamma_\ell > \max_{i\in\mcS}\gamma_i \left| \gamma_\ell=\max_{k\in\mcB}\gamma_k \right. \right] \bbP\left[  \gamma_\ell=\max_{k\in\mcB}\gamma_k  \right]
\label{eq:additivity_proof2}
\\ & = \sum_{\ell\in\mcB\setminus\mcS} \bbP\left[  \gamma_\ell=\max_{k\in\mcB}\gamma_k  \right]
\label{eq:additivity_proof3}
\\ & = 1 - \sum_{\ell\in\mcS} \bbP\left[  \gamma_\ell=\max_{k\in\mcB}\gamma_k  \right],
\label{eq:additivity_proof4}
\end{align}
where \eqref{eq:additivity_proof2} is the application of the law of total probability on the event $\left\{ \gamma_\ell=\max_{k\in\mcB}\gamma_k \right\}$, and \eqref{eq:additivity_proof3} follows because if $\ell\in\mcS$ then $\bbP\left[ \gamma_\ell > \max_{i\in\mcS}\gamma_i \left| \gamma_\ell=\max_{k\in\mcB}\gamma_k \right. \right]=0$ and if $\ell\in\mcB\setminus\mcS$ then $\bbP\left[ \gamma_\ell > \max_{i\in\mcS}\gamma_i \left| \gamma_\ell=\max_{k\in\mcB}\gamma_k \right. \right]=1$. Using the fact that $\sum_{\ell\in\mcB} \bbP\left[  \gamma_\ell=\max_{k\in\mcB}\gamma_k  \right]=1$, we obtain \eqref{eq:additivity_proof4}. From \eqref{eq:additivity_proof4}, we have 
\begin{align}
\bar{P}_\rmpl(1,\mcS) = \sum_{\ell\in\mcS} \bbP\left[  \gamma_\ell=\max_{k\in\mcB}\gamma_k  \right].
\end{align}
Thus, for any $\mcS\subset\mcT\subset\mcB$ and $\forall n\in\mcB\setminus\mcT $, we have $\bar{P}_\rmpl(1,\mcS\cup\{n\})-\bar{P}_\rmpl(1,\mcS)=\bar{P}_\rmpl(1,\mcT\cup\{n\})-\bar{P}_\rmpl(1,\mcT)=\bbP\left[  \gamma_n=\max_{k\in\mcB}\gamma_k  \right]$, which is the definition of modular functions \cite{Nemhauser1978,Calinescu2011}. 
\end{IEEEproof}

While the solution in Algorithm \ref{alg:min_power_loss_algorithm} is intuitive, 
using a brute force search to solve the minimization problem at each selection step is not efficient. 
At each selection step, we need to evaluate the power loss probability $|\mcB\setminus\mcS_{n-1}|\le |\mcB|-N_\rmb$ times. Since $|\mcB|$ typically is much larger than $N_\rmb$, this means that the total number of evaluations is $\mcO(N_\rmb|\mcB|)$. 
Defining the probability of being optimal as
\begin{align}
P_\mathrm{opt}(i) & = \bbP\left[  \gamma_i=\max_{k\in\mcB}\gamma_k  \right]
\\ & = \bbP\left[ \gamma_i \ge \gamma_k, \forall k\in\mcB\setminus\{i\} \right],
\end{align} 
the proof of Proposition \ref{prop:modularity} suggests a more efficient solution. From \eqref{eq:additivity_proof4}, we see that minimizing $P_\rmpl(1,\mcS_{n-1}\cup\{i\})$ over $i\in\mcB\setminus\mcS_{n-1}$ is equivalent to solving 
\begin{align}
k=\underset{i\in\mcB\setminus\mcS_{n-1}}{\arg\max}P_\mathrm{opt}(i).
\end{align}
This means that Algorithm \ref{alg:min_power_loss_algorithm} is equivalent to selecting the beam pairs by ranking their probability of being optimal in descending order. This solution requires to compute the probability of being optimal $\mcO(|\mcB|)$ times. 

\begin{algorithm}
 \caption{Greedy candidate beam pair selection}
 \label{alg:min_power_loss_algorithm}
 \begin{algorithmic}[1]
  \STATE $\mcS_0 \gets \emptyset$
  \FOR {$n=1:N_\rmb$}
  \STATE {$\mcS_n \gets \mcS_{n-1} \cup \underset{i\in\mcB\setminus \mcS_{n-1}}{\arg\min}\, P_\rmpl(1,\mcS_{n-1}\cup\{i\})$}
  \ENDFOR
 \end{algorithmic} 
\end{algorithm}

We now present how to compute $P_\mathrm{opt}(i)$. Note that Type B cannot be used to compute $P_\mathrm{opt}(i)$ because it only stores the average received powers. Denote $\gamma_{nk}$ the received power observed at the $k$-th beam pair in the $n$-th observation, $P_\mathrm{opt}(i)$ is estimated using Type A fingerprints by 
\begin{align}
\label{eq:prob_opt_typeA}
P_\mathrm{opt}(i) \simeq \frac{1}{N}\sum_{n=1}^{N} \bone\left( \gamma_{ni} > \gamma_{nk}, \forall k\in\mcB\setminus\{i\} \right),
\end{align}
where $N$ is the number of observations in the fingerprint database (number of rows of Table~\ref{tab:raw_fingerprint_example}), and $\bone(E)$ is the indicator function which outputs 1 if $E$ is true and 0 otherwise.
Note that if we choose to keep $M < |\mcB|$ measurements per contributing vehicle, 
not all $\gamma_{nk}$ for $k=1,2,\dots,|\mcB|$ are recorded. We assume those $\gamma_{nk}$ that are not recorded to be zero in \eqref{eq:prob_opt_typeA}. This is a reasonable approximation because these $\gamma_{nk}$ are much smaller than the received power of the top ranked beam pairs that are recorded in the database. 
The expression in \eqref{eq:prob_opt_typeA} is equivalent to counting how often the $i$-th beam pair is observed to be the strongest. 
This means that $P_\mathrm{opt}(i)$ is estimated to be 0 for all beam pairs that have not yet been seen to be the strongest in the database collected. Less important beam pairs that rarely provide the strongest received power are difficult to rank using \eqref{eq:prob_opt_typeA} with a reasonable $N$. 
In fact, in our simulation in Section \ref{sec:compare_beam_selection} with $N=450$, there are about 30 distinct beam pairs that are observed to be the best at least once in the database. This means that using \eqref{eq:prob_opt_typeA}, we can produce a ranked list of length up to around 30. Since this list is short, it could happen that the allowable training budget $N_\rmb$ is larger than 30, and we want to produce a longer list that ranks the less important beam pairs while using the same database. 
We propose to rank these less important beam pairs by the same metric but computed while ignoring the correlation in the fingerprints. By assuming the independence between the pairs, we have
\begin{align}
P_\mathrm{opt}(i) & = \bbE_{\gamma_i}\left[ \prod_{k\in\mcB\setminus\{i\}} \bbP \left[ \gamma_i > \gamma_k | \gamma_i \right] \right] 
\label{eq:opt_probB1}
\\& \simeq \frac{1}{N}\sum_{n=1}^{N} \prod_{k\in\mcB\setminus\{i\}} \frac{1}{N} \sum_{m=1}^{N} \bone\left( \gamma_{ni} > \gamma_{mk} \right).
\label{eq:opt_probB2}
\end{align} 
To summarize, this beam pair selection method uses both \eqref{eq:prob_opt_typeA} and \eqref{eq:opt_probB2} to produce a ranked list of beam pairs. 
Let $\mcB_\mathrm{nz}$ be the set of all beam pairs with nonzero $P_\mathrm{opt}(i)$ according to \eqref{eq:prob_opt_typeA}, then the top-$|\mcB_\mathrm{nz}|$ in the ranked list are obtained using \eqref{eq:prob_opt_typeA}, and the rest of the beam pairs $\mcB\setminus \mcB_\mathrm{nz}$ are ranked using \eqref{eq:opt_probB2}.   
We call this beam pair selection method MinMisProb.

\section{Numerical Results and Discussions} \label{sec:numerical_result}
This section provides numerical evaluations of the proposed beam alignment. 
All evaluations here use a dataset of 500 channel samples generated from the ray-tracing simulator (see Fig. \ref{fig:WI_sim_scenario}) and assume $16\times16$ UPAs at both the CV and RSU unless stated otherwise. The codebook for the $16\times16$ array has 271 beams. 
We conduct $K$-fold cross validation, with $K=10$ as recommended in \cite{book-elements-statistical-learning}. Specifically, the dataset of 500 channel samples is divided into 10 subsets (or folds) of size 50 each. 
Then, nine of them are used as the training set to build the database, and the remaining one is used as the test set to evaluate the proposed beam alignment. This is repeated 10 times, where each time a different subset is selected as the test data. When a different evaluation method is used, it will be stated explicitly. 
Common simulation parameters are summarized in Table \ref{tab:sim_params}. 

\begin{table}
\centering
\caption{Common simulation parameters}
\label{tab:sim_params}
\begin{tabular}{|c|rl|} \hline 
Parameters & \multicolumn{2}{c|}{Value} \\ \hline\hline
Carrier frequency & 60 \hspace{-0.5cm} & \hspace{-0.0cm} GHz \\ \hline
Bandwidth & 1760 \hspace{-0.5cm} & \hspace{-0.0cm} MHz \\ \hline
Antenna array & 
16$\times$16 \hspace{-0.5cm} & \hspace{-0.0cm} UPA \\ \hline
Mean vehicle gap & \hspace{0.2cm}4.78 \hspace{-0.5cm} & \hspace{-0.0cm} m \\ \hline
\end{tabular}
\end{table}

The two types of database are built in the following manners in the simulations. 
For Type A fingerprints, each contributing vehicle conducts a full exhaustive beam measurements and the top-100 beam pairs are recorded as explained in Section \ref{sec:multipath_fingerprint}. 
For a fair comparison, the two types of database should be built using the same number of measurement data. Thus, the database for Type B is obtained by summarizing Type A database. Specifically, instead of keeping all the raw received powers, only the average received power is recorded for each beam pair. We note that in actual implementations of Type B data collection, a full exhaustive search can be collected by a number of vehicles depending on the time budget the vehicles have as mentioned in Section \ref{sec:multipath_fingerprint}. 


\subsection{Performance Comparison of Proposed Beam Pair Selection Methods} \label{sec:compare_beam_selection}
This subsection presents a performance comparison of the proposed beam alignment method with the two candidate beam pair selection methods, namely AvgPow and MinMisProb, when the measurement noise is negligible. The impact of noise will be considered in the next subsection. 

\begin{figure} 
\centering
\includegraphics[width=\figwidthscale\columnwidth]{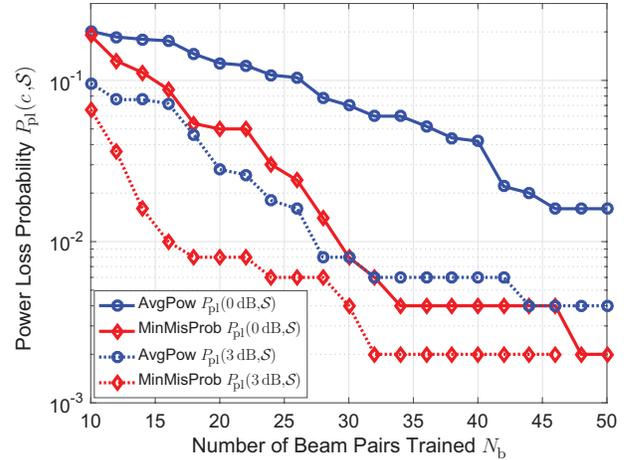}
\caption{Power loss probability versus the number of beam pairs trained. The 3 dB power loss probability plot ends before reaching $N_\rmb=50$ because there is no such instance of power loss computed from the cross validation. MinMisProb outperforms AvgPow in both the misalignment and 3 dB power loss probability.   
}
\label{fig:powLoss_prob_vs_Nb}
\end{figure}

Fig. \ref{fig:powLoss_prob_vs_Nb} compares the two beam pair selection methods in terms of the power loss probability. 
Two different levels of power loss severity are shown: the misalignment probability $P_\rmpl(\text{0 dB},\mcS)$ and the probability that the power loss is less than 3 dB $P_\rmpl(\text{3 dB},\mcS)$ (called the 3 dB power loss probability). 
MinMisProb dominates AvgPow in both levels of misalignment. 
This is expected because MinMisProb is optimal (in terms of the misalignment probability) by its definition that exploits the correlation between the different beam pairs available in Type A fingerprints. 
We note that when computing the probability of being optimal $P_\mathrm{opt}(\cdot)$ using \eqref{eq:prob_opt_typeA}, the number of beam pairs with nonzero $P_\mathrm{opt}(\cdot)$ is around 30 (the exact number depends on the folds chosen for training). 
The plots of MinMisProb become flat at $N_\rmb$ of around $30$. This means that the complementary selection using \eqref{eq:opt_probB2} does not perform as well as when using \eqref{eq:prob_opt_typeA} that exploits the correlation information. It, however, can still identify relevant beam pairs without additional training data. 
From these results, we conclude that if Type A fingerprints are available, MinMisProb is the choice; otherwise, the AvgPow method should be used. 

\subsection{Required Training Sample Size} \label{sec:req_sample_size}
This subsection provides an empirical evaluation to estimate the training sample size to build the fingerprint database. By sample size here, we mean the number of exhaustive beam measurements conducted to collect the data (i.e., the number of rows of Table \ref{tab:raw_fingerprint_example}). We start with the description of the evaluation method. We still use the 10-fold cross validation as before, but now instead of using all the nine folds (450 samples) for training, we only use a subset of $N<450$ of these samples. To average out the dependence on the sampling of the subset, we repeat the evaluation of the test set 50 times, where in each time we randomly choose $N$ samples out of the available training set of $450$ samples to build the database. Both AvgPow and MinMisProb show a similar trend, and we show only the results for AvgPow here. 

\begin{figure} 
\centering
\subfigure[Average 3 dB power loss probability]{%
\includegraphics[width=\figwidthscale\columnwidth]{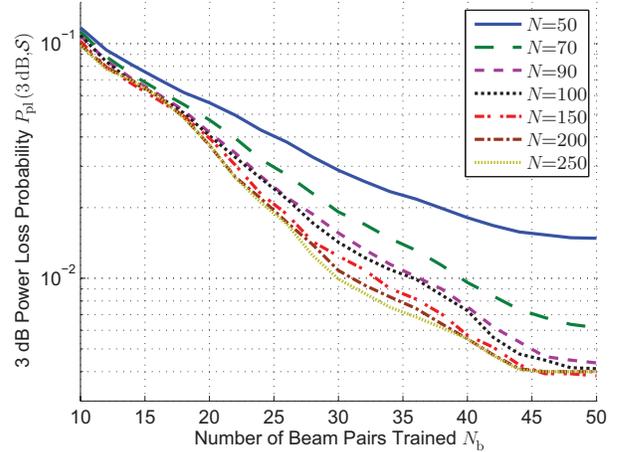}
\label{fig:3dB_powLoss_prob_diff_training_sample_size_AvgPow_semilogy}
}
\quad
\subfigure[Learning curve at $N_\rmb=30$]{%
\includegraphics[width=\figwidthscale\columnwidth]{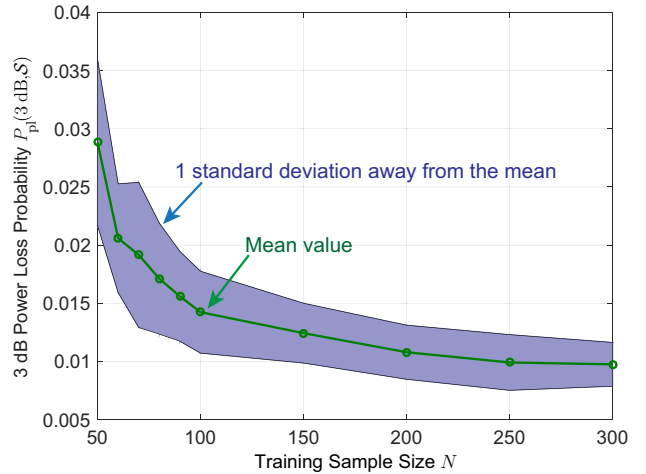}
\label{fig:Learning_curve_3dB_powLoss_prob_AvgPow_WB}
}
\caption{3 dB power loss probability of AvgPow as a function of the training sample size $N$. Fig. \ref{fig:3dB_powLoss_prob_diff_training_sample_size_AvgPow_semilogy} shows the average 3 dB power loss probability for different $N$. Fig. \ref{fig:Learning_curve_3dB_powLoss_prob_AvgPow_WB} shows the learning curve in terms of the 3 dB power loss probability when the number of beam pairs measured is set to $N_\rmb=30$. We can see from the plots that the improvement diminishes for subsequent increase in the training sample size. 
}
\label{fig:learning_curve}
\end{figure}

We evaluate the quality of the fingerprint obtained using the training set of size $N$ by the average of the power loss probabilities estimated by the 50 cross validations as described earlier. 
Fig. \ref{fig:learning_curve} shows average 3 dB power loss probabilities for the training sample size ranging from 50 to 250. We can see a large improvement when increasing $N$ from 50 to 90. Subsequent increases in $N$, however, provide diminishing improvement. To see this effect more clearly, we plot in Fig. \ref{fig:Learning_curve_3dB_powLoss_prob_AvgPow_WB} the 3 dB power loss probability when the number of beam pairs trained is fixed at $N_\rmb=30$. This plot is typically known as the learning curve \cite{book-elements-statistical-learning}, which quantifies the improvement as the training sample size (i.e., the learning effort) increases. The figure shows the mean and the region of one standard deviation from the mean. We see a sharp improvement up to around $N=100$, and a slower improvement beyond that. 
We thus conclude that a training sample size of around 200 to 250 seems good enough. 

\begin{figure}
\centering
\includegraphics[width=\figwidthscale\columnwidth]{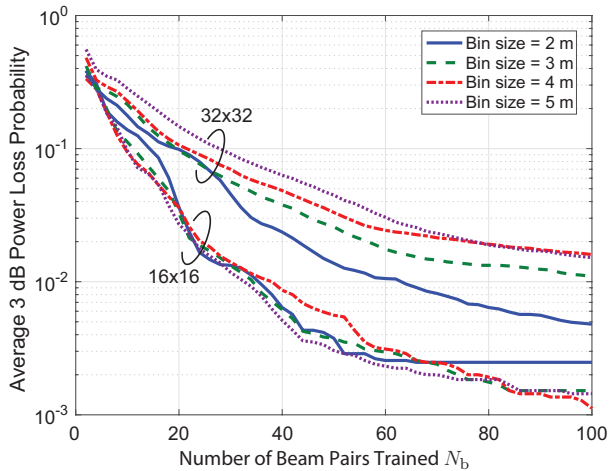}
\caption{Average 3 dB power loss probability when using different location bin sizes. All bin sizes performs similarly when using $16\times16$ arrays. When using the larger $32\time32$ arrays, smaller bin sizes performs better. 
}
\label{fig:3dBPowLossProb_AvgPow_diff_bin_sizes}
\end{figure}
\subsection{Effect of Location Bin Size} \label{sec:effect_bin_size}
We start with the description of the evaluation method. We use 10-fold cross validations on 300 channel samples where the CV is within the location bin. Recall from Section \ref{sec:channel_model} that the channel samples are generated with the center of the location bin at $d_0=30$ m and the CV is randomly placed in the range $[d_0-2.5,d_0+2.5]$. For example, when evaluating the performance for a bin size of 2 m, we only use the channel samples where the CV's center position (the position of its antenna) is within $[d_0-1,d_0+1]$. 
Since each cross validation is computed on a different dataset, we need to perform averaging to eliminate dependency on the dataset. To do this, we repeat the cross validation 50 times, where each time 300 channel samples are randomly selected from the set of the channel samples where the CV is within the location bin. To be able to evaluate small bin sizes, we generated more channel samples (total 1000 samples) to ensure that we have enough samples for the averaging. 
For the comparison metric, we use the average 3 dB power loss probability, which is obtained by averaging over the results from the 50 cross validations. 
AvgPow is used as the beam selection method.

Fig. \ref{fig:3dBPowLossProb_AvgPow_diff_bin_sizes} shows the average 3 dB power loss probability for location bin sizes of 2, 3, 4, and 5 m when using UPA $16\times16$ and $32\times32$. 
The variation of the bin size from 2 m to 5 m has little impact when using the $16\times16$ arrays. We note that the plots are in log-scale and the gap at $N_\rmb=100$ is small (it is less than 0.002). When using a larger array of $32\times32$, which requires higher location precision, we can see that a smaller bin size provides better performance. 
The physical reason why the performance is not that sensitive to the location bin size is that NLOS paths are less affected by the position accuracy than a LOS path. This is because NLOS paths have nonzero angular spread which makes it easier for beam alignment. Instead of having to align to a single direction as in the LOS path, the beam can be aligned to within the range of the angular spread. 
From these results, we can conclude that location bin size of 5 m is good enough when using UPA $16\times16$. When using a large array such as $32\times32$, smaller bin sizes provide better performance. 
Finally, note that while smaller bin sizes are preferred for beam alignment performance, it has to be large enough to account for the level of position accuracy available to the vehicles. 

\subsection{Effect of Measurement Noise} \label{sec:numerical_result_noise_impact}

\begin{figure}
\centering
\includegraphics[width=\figwidthscale\columnwidth]{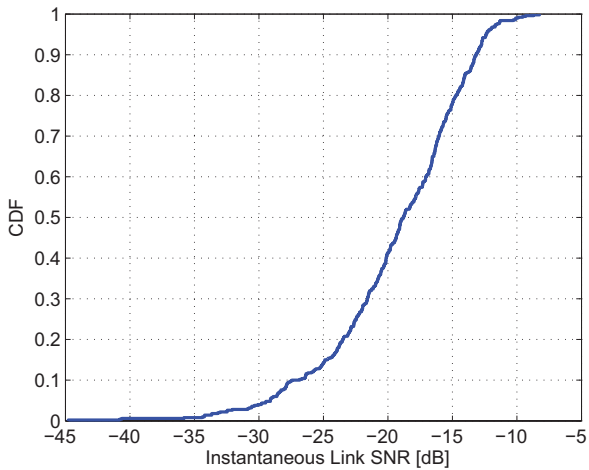}
\caption{CDF of the link SNR of the generated channels.}
\label{fig:cdf_instantaneous_link_SNR}
\end{figure}

This subsection studies the effect of measurement noise on the beam alignment accuracy. The results are shown in terms of the Equivalent Isotropic Radiated Power (EIRP), which is the transmit power plus the transmit antenna gain (in dB scale). EIRP is used instead of the transmit power because it is regulated \cite{mmwave-vehicular-survey}. 
To provide the context of the operating SNR, we start with a description of the link SNR of the channel samples generated from the ray-tracing simulation. 
We assume the noise power is given by $P_\rmn=-174+10\log_{10}B$ dBm, where $B=1760$ MHz is the sampling rate defined in IEEE 802.11ad for single carrier PHY \cite{802.11ad}. Denote $P_0$ the received power when isotropic antennas (0 dBi antenna gain) are used at both the transmitter and the receiver with $0$ dBm transmit power, the link SNR is defined as 
\begin{align}
\SNR = \frac{P_0}{P_\rmn}.
\end{align}
Fig. \ref{fig:cdf_instantaneous_link_SNR} shows the CDF of the link SNR computed from the received powers output from the ray-tracing simulation. 
We note that with an EIRP of 0 dBm, the SNR at the receiver (before the receive antenna gain) is the link SNR. The average link SNR is $-16.0$ dB, and thus the average SNR at the receiver is around 0 dB when using an EIRP of 16 dBm.


\begin{figure} 
\centering
\subfigure[Misalignment probability with noise]{%
\includegraphics[width=\figwidthscale\columnwidth]{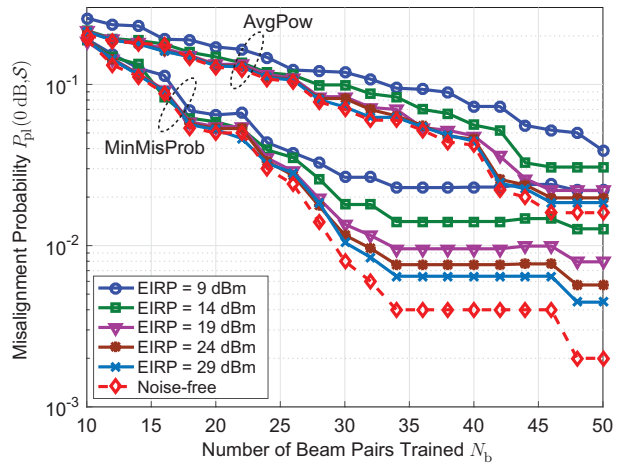}
\label{fig:misalignment_prob_vs_Nb_withNoise}
}
\quad
\subfigure[3 dB power loss probability with noise]{%
\includegraphics[width=\figwidthscale\columnwidth]{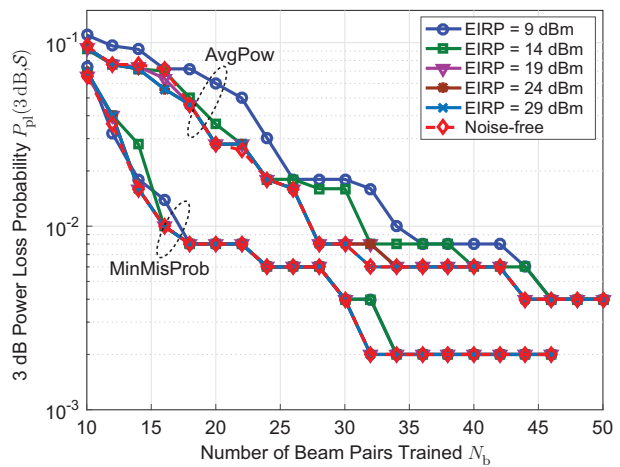}
\label{fig:3dB_powLoss_prob_vs_Nb_withNoise}
}
\caption{Power loss probability as a function of $N_\rmb$ in the presence of noise. The noise impacts the misalignment probability much more than the 3 dB power loss probability. 
}
\label{fig:power_loss_probability_with_noise}
\end{figure}

Fig. \ref{fig:power_loss_probability_with_noise} shows a comparison of the power loss probability with and without noise for the two beam selection methods. It is assumed that the same EIRP is used during the database collection and when doing beam measurements for beam alignment. We can see that the misalignment probability is much more affected than the 3 dB power loss probability. This is because for a 3 dB power loss event to happen the noise must be large enough to flip the order of the optimal pair with a beam pair that has the power of at least 3 dB below the optimal beam pair, which occurs much less frequent than the misalignment event (i.e., any nonzero power loss). We note that the plots are in log-scale, and the gaps in the misalignment probability between the EIRP$=9$ dBm case and the noise-free case are 0.03 and 0.02 at $N_\rmb=50$ for AvgPow and MinMisProb, respectively. Thus, overall we can conclude that MinMisProb is less affected by noise than AvgPow. 

\begin{figure} 
\centering
\includegraphics[width=\figwidthscale\columnwidth]{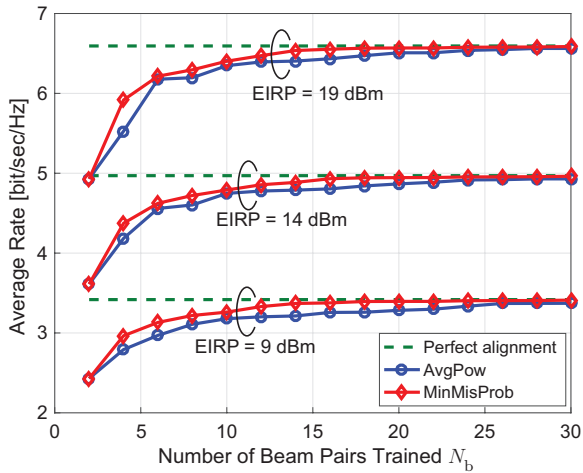}
\caption{Average rate of the proposed beam alignment compared to the perfect alignment case. MinMisProb consistently achieves a higher average rate than AvgPow for the same $N_\rmb$. The rate loss becomes negligible at $N_\rmb$ at around 20 and 30 for MinMisProb and AvgPow, respectively.  
}
\label{fig:rate_plot}
\end{figure}

We show the average rates when using the proposed beam alignment in Fig. \ref{fig:rate_plot}.  
The instantaneous rate is computed using  
\begin{align}
R_\mathrm{ins}=\log_2\left( 1 + \frac{ P_\rmt\|\bh_s\|^2 }{P_\rmn} \right),
\end{align}
where $\bh_s$ is the effective channel of the beam pair selected after the beam training. 
Fig. \ref{fig:rate_plot} shows the average rate as a function of the number of beam pairs trained $N_\rmb$ for EIRP of 9, 14, and 19 dBm. Increasing the training overhead $N_\rmb$ improves the alignment quality leading to higher average rates. 
The rate loss compared to the perfect alignment case becomes negligible at around $N_\rmb=20$ for MinMisProb and at around $N_\rmb=30$ for AvgPow. 
The gaps to the perfect alignment at small $N_\rmb$ do not improve with increasing EIRP. This is due to the larger power loss probability when using a small $N_\rmb$ and so cannot be eliminated by increasing the transmit power. 


\subsection{Effect of Traffic Mismatch during Database Collection and Exploitation} \label{sec:sensitivy_to_traffic_density}
In this subsection, we provide some simulation results to show the effect of the mismatch of the traffic density during the database collection and exploitation. We expect the effect to be more pronounced when the difference in traffic density is large. We, thus, study the high and low traffic density cases. 
To do this, we generated another dataset of 500 channel samples using the ray-tracing simulation with the same setting as described in Section \ref{sec:channel_model} but with a lower vehicular traffic using $\mu_\zeta=0.0536$ (average vehicle gap of 18.66 m) and the car-to-truck ratio of 9:1. We use all combinations of these two datasets as the training and test set to evaluate the performance of the proposed beam alignment, namely the four combinations of training and test set of (low,low), (high,low), (low,high), and (high,high). We note that (high,high) is what is used so far. AvgPow is used as the selection method here. 

\begin{figure} 
\centering
\subfigure[3 dB power loss probability.]{%
\includegraphics[width=\figwidthscale\columnwidth]{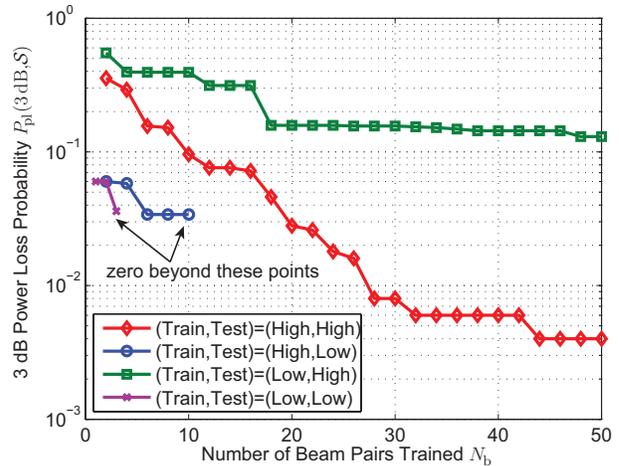}
\label{fig:3dB_powLoss_prob_vs_Nb_diff_traffic} }
\quad 
\subfigure[Normalized averaged rate when EIRP=24 dBm.]{%
\includegraphics[width=\figwidthscale\columnwidth]{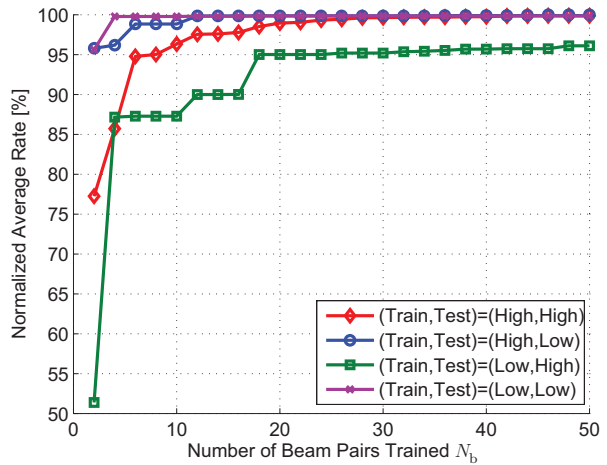}
\label{fig:rate_vs_number_of_training_Nb_EIRP=24dBm_16x16_diff_traffic}
}
\caption{ 
Effect of the mismatch in traffic density during database collection and exploitation. Database collected in a light traffic does not work well when used in a dense traffic because the database cannot captures all the paths in the richer scattering environment of the dense traffic. On the contrary, database collected in a dense traffic works well regardless of the traffic condition when it is exploited. It only has a slightly degraded efficiency when used in a low traffic condition. 
}

\label{fig:sensitivity_to_traffic_conditions}
\end{figure}

Fig. \ref{fig:sensitivity_to_traffic_conditions} shows the performance in terms of the 3 dB power loss probability and the average rate normalized by the perfect beam alignment case when using EIRP of 24 dBm. When exploiting in the low traffic setting, training (i.e., building the fingerprint database) in either the low or high traffic condition yields good performance while training in the low traffic density (i.e., matched traffic condition) is slightly more efficient. On the contrary, when exploiting in the high traffic condition, 
the performance loss due to the mismatch in the traffic conditions during database collection and exploitation is higher. 
Intuitively, this is because the database collected under a low traffic density cannot adequately capture paths in the richer scattering environment of the high traffic condition. The same trend can be confirmed in the rate plot in Fig. \ref{fig:rate_vs_number_of_training_Nb_EIRP=24dBm_16x16_diff_traffic}. We note that the largest loss observed is around 5\% when using the database collected in the low traffic condition and used in the high traffic case. 
These results show that building the database in the same traffic condition as when the database is used provides the best performance. If adaptation to the traffic condition is not possible or costly, 
then the database should be collected in high traffic conditions.

\subsection{Comparison with Existing Techniques} \label{sec:numerical_comparison}


This subsection compares the performance of the proposed method with two existing solutions. The first one is a hierarchical beam search adopted in IEEE 802.11ad, and the second one uses only the position to determine the beam pointing direction.

We start with a comparison with the IEEE 802.11ad method. 
IEEE 802.11ad beam alignment is a beam sweeping method using a hierarchical beam codebook structure to reduce the amount of beam training \cite{802.11ad,Hosoya:MIDC-a-novel-beamforming-technique:14}. 
It is required by the standard that the antenna gain of the quasi-omni pattern (the widest level in the codebook) be at most 15 dB ($\simeq32$ in linear scale) lower than a directional pattern \cite[Section 21.10.1]{802.11ad}. Because of this constraint, we consider a two-level beam codebook: the quasi-omni and the sector level. We further assume for simplicity that the number of codewords at the sector level is equal to $N_\rma$, the number of elements of the array (e.g., when using a 2D DFT codebook). Since the gain in the main beam direction of an array is $N_\rma$, we have $N_\mathrm{sec}=N_\rma$ and $N_\mathrm{QO}=N_\mathrm{sec}/32$, where $N_{\mathrm{sec}}$ is the number of sector beams and $N_\mathrm{QO}$ is the number of quasi-omni patterns. Since the quasi-omni patterns are the widest in the codebook, an exhaustive search is needed at the quasi-omni level to determine the best quasi-omni pair. 
Once the best quasi-omni pair has been identified, IEEE 802.11ad uses a low complexity single-sided search to find the best receive and transmit sectors. A single-sided search is conducted 
by having the transmitter use the best transmit quasi-omni pattern while the receiver sweeps over the sectors whose pointing directions are within the best receive quasi-omni pattern. The same procedure is used to find the best transmit sector.
The total beam training time (excluding feedbacks) is given by
\begin{align}
T_\mathrm{11ad} = N_\mathrm{QO}^2 T_\mathrm{QO} + 2 \frac{N_\mathrm{sec}}{N_\mathrm{QO}} T_\mathrm{sec},
\label{eq:11ad_training_duration}
\end{align}
where $T_\mathrm{QO}$ and $T_\mathrm{sec}$ are the duration of a training frame at the quasi-omni and the sector level, respectively. The second term in \eqref{eq:11ad_training_duration} is based on the assumption that each quasi-omni pattern covers the same number of sectors. The factor two is because the single-sided search has to be conducted for both the transmitter and the receiver. We note that quasi-omni patterns have low antenna gain and thus require a large spreading factor to compensate for the lack of antenna gain. IEEE 802.11ad uses 32$\times$ spreading for this. A beam training at the quasi-omni level is done by sending an SSW (sector sweep) frame of length $26.8$ $\mu$s, which consists of $4.3$ $\mu$s for preamble and $22.5$ $\mu$s for header and information in the SSW frame. Since the preamble might be needed for synchronization purpose, we assume that it is unchanged and set $T_\mathrm{sec}=4.3+22.5/32=5.0$~$\mu$s.   

\begin{figure}
\centering
\includegraphics[width=\figwidthscale\columnwidth]{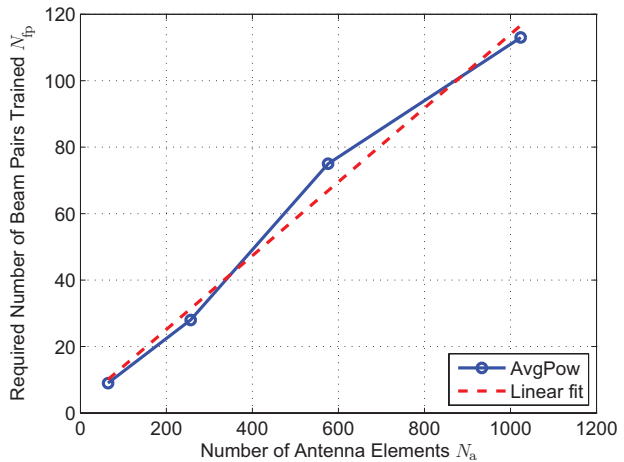}
\caption{Required amount of beam training of the proposed method. The overhead increases roughly linearly with the number of antenna elements of the array.}
\label{fig:req_num_beam_trainings_vs_num_ante_element_avgPow}
\end{figure}

We now compute the overhead of the proposed approach. As discussed in Section \ref{sec:multipath_fingerprint}, the fingerprint database is expected to be valid for a long period of time and thus the cost per usage after the database is collected will be negligible. We, therefore, consider only the beam training overhead here. 
We define the overhead as the number of beam pairs trained $N_\mathrm{fp}$ needed to achieve $P_\rmpl(\mathrm{3\,dB},\mcS)\le 1$\%. Since our approach does not use wide beams for the beam training, we assume the training duration to be $T_\mathrm{sec}$, and the total training time is
\begin{align}
T_\mathrm{fp} = N_\mathrm{fp}T_\mathrm{sec}.
\end{align}
Fig. \ref{fig:req_num_beam_trainings_vs_num_ante_element_avgPow} shows the required amount of beam training $N_\mathrm{fp}$ for UPAs of sizes $8\times8$, $16\times16$, $24\times24$, and $32\times32$ when using AvgPow as the beam pair selection method. The codebook sizes are 87, 271, 641, and 1047, respectively. 
The plot shows $N_\mathrm{fp}$ as a function of the number of elements $N_\mathrm{a}$, which shows a roughly linear trend in $N_\rma$. 

\begin{figure}
\centering
\includegraphics[width=0.7\columnwidth]{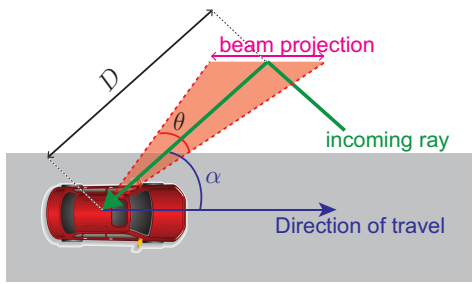}
\caption{An illustration of the beam coherence time concept. For the beam to stay aligned, the reflection point must be within the beam projection. The beam coherence time is the average time that the reflection point is within the beam projection.}
\label{fig:illustration_beam_coherence_time}
\end{figure}

To understand this overhead in the mobility context, we leverage the concept of beam coherence time \cite{Va:Impact-of-beamwidth:2016}, which is the duration before beam realignment is required. The beam coherence time is the duration that the pointing error due to mobility causes the received power to drop by some threshold from the peak. Since the codebook quantizes the angular domain by the 3 dB beamwidth, the initial pointing error ranges in $[-\Theta/2,\Theta/2]$, where $\Theta$ is the 3 dB beamwidth. Assuming the initial pointing error to be uniform in $[-\Theta/2,\Theta/2]$, then the beam coherence time is the duration that the pointing error becomes larger than half the 3 dB beamwidth. Using the pointing error derived in \cite{Va:Impact-of-beamwidth:2016}, the beam coherence time $T_\rmB$ can be written as 
\begin{align}
T_\mathrm{B} = \frac{D}{v\sin\alpha} \frac{\Theta}{2},
\end{align}
where $D$ is the distance to the reflector/scatter, $\alpha$ is the main beam direction with respect to the direction of travel, and $v$ is the speed of the CV (see Fig. \ref{fig:illustration_beam_coherence_time}). We note that $D$ refers to the distance from the receiver to the nearest reflector/scatter, and this concept can be applied to channels with high orders of reflections (although, paths undergoing multiple reflections are typically weak at mmWave frequencies). 
We note that this result is based on a 2D model considering only the azimuth. This, however, is applicable here because there is negligible change in the elevation angle as the vehicle moves. Since we are considering a square array with a total number of elements of $N_\rma$, the azimuth beamwidth $\Theta$ can be approximated by the beamwidth of a uniform linear array of size $\sqrt{N_\rma}$ given by $\Theta\simeq 0.886\frac{2}{\sqrt{N_\rma}}$ when the antenna spacing is set to half the wavelength \cite[p. 885]{Orfanidis2014}. Using this approximation, we have
\begin{align}
T_\mathrm{B}(N_\rma) = \frac{0.886 D}{v\sqrt{N_\rma}\sin\alpha} .
\end{align}
From the geometry, we set $\alpha=60^\circ$ and $D=12$ m, which represents a typical reflection path off the buildings in Fig. \ref{fig:WI_sim_scenario}. Note that these parameters are chosen as a representative worst case example with a reasonably small distance to the nearest reflector (a building in this case) and reasonably large $\alpha$. The worst value possible for $\alpha$ is $90^\circ$, but this path direction is unlikely in the geometry shown in Fig. \ref{fig:WI_sim_scenario}. 

\begin{figure} 
\centering
\subfigure[Beam training duration versus the number of antenna elements.]{%
\includegraphics[width=\figwidthscale\columnwidth]{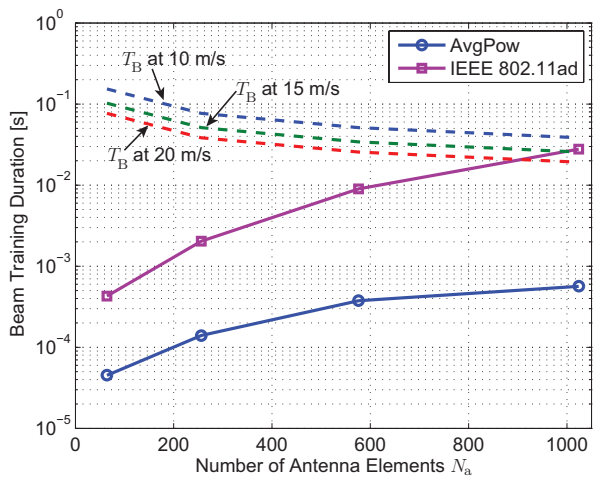}
\label{fig:total_beam_training_duration} }
\quad 
\subfigure[Average rate considering beam training overhead (transmit power of 0 dBm).]{%
\includegraphics[width=\figwidthscale\columnwidth]{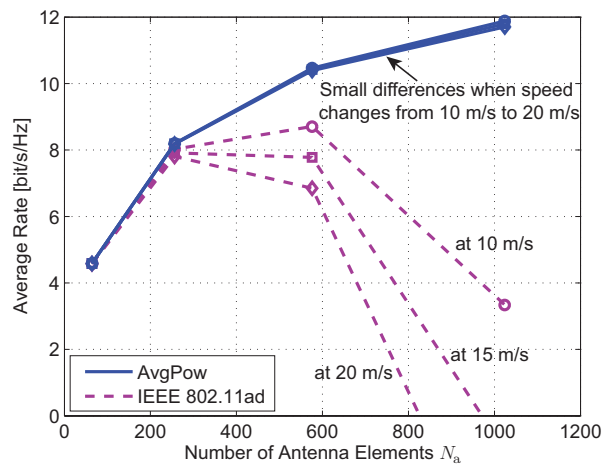}
\label{fig:Average_rate_considering_overhead_and T_B}
}
\caption{Overhead comparison between the proposed method and that of IEEE 802.11ad. The beam training time of the proposed method only takes up to a few percent of the beam coherence time $T_\rmB$, while that of IEEE 802.11ad can exceed $T_\rmB$ when using a large array leaving no time for data communications. 
}
\label{fig:rate_plot_with_overhead}
\end{figure}

Fig. \ref{fig:rate_plot_with_overhead} compares the overhead of the proposed beam alignment using AvgPow selection method with that of IEEE 802.11ad. Fig. \ref{fig:total_beam_training_duration} shows beam training durations as a function of the array size $N_\rma$. We recall that this beam training duration does not include the initial training request and feedback, which do not depend on the array size. We also plot the beam coherence times $T_\rmB$ when the vehicle speed is 10, 15, and 20 m/s. The training duration of the proposed method is at most a few percents of $T_\rmB$, while that of IEEE 802.11ad can exceed $T_\rmB$ when the array becomes large enough. This means that IEEE 802.11ad cannot finish the training before realignment is required. Fig. \ref{fig:Average_rate_considering_overhead_and T_B} compares the average rate when taking the beam training duration and $T_\rmB$ into account assuming a transmit power of 0 dBm (corresponding to $24$ dBm EIRP when using a $16\times16$ array). The average rate here is defined as 
\begin{align}
\label{eq:avg_rate_including_beam_training_overhead}
R_\mathrm{avg} = \frac{T_\rmB - T_\mathrm{trn}}{T_\rmB} R_\mathrm{trn},
\end{align}
where $T_\mathrm{trn}$ is the beam training duration and $R_\mathrm{trn}$ is the average rate after the alignment. For the proposed method, $R_\mathrm{trn}$ is determined from rate plots when using the different array sizes (Fig. \ref{fig:rate_plot} shows rate plots when using $16\times 16$ arrays). We assume optimistically that IEEE 802.11ad achieves the perfect alignment rate. While the average rate of the proposed beam alignment keeps increasing as the array size increases, that of IEEE 802.11ad increases slowly at 10 m/s or starts to decrease at speed beyond 15 m/s when the array becomes larger than $16\times 16$. We also observe that the average rate of IEEE 802.11ad becomes zero at $N_\rma$ around 800 and 1000 with when the speed is 15 and 20 m/s, respectively. This is because the training duration becomes larger than $T_\rmB$, and there is no time left for data communication.

\begin{figure} 
\centering
\subfigure[High traffic density.]{%
\includegraphics[width=\figwidthscale\columnwidth]{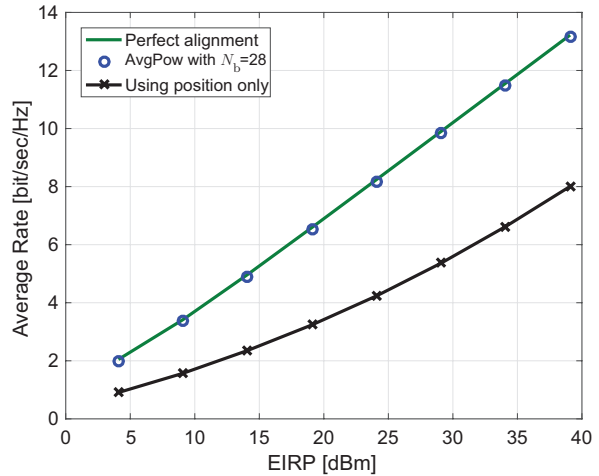}
\label{fig:avg_rate_comparison_posi_only_high_traffic} }
\quad 
\subfigure[Low traffic density.]{%
\includegraphics[width=\figwidthscale\columnwidth]{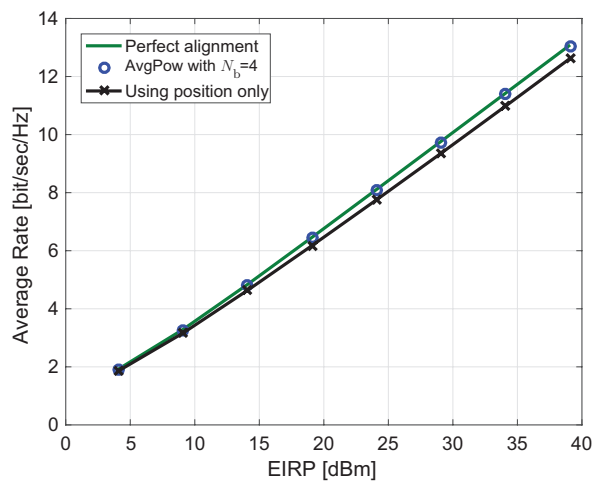}
\label{fig:avg_rate_comparison_posi_only_low_traffic}
}
\caption{Average rate comparison with the beam alignment method using position only. The method using only the position does not need beam training but works only when the LOS path is available. It performs well in a low traffic density where LOS path is often available but performs poorly in a dense traffic where the LOS path is often blocked. Taking advantage of the fingerprint database, the proposed method works well in both cases.   
}
\label{fig:avg_rate_comparison_posi_only_diff_traffic}
\end{figure}

We now compare the performance of the proposed beam alignment with the method that uses position only. Fig. \ref{fig:avg_rate_comparison_posi_only_diff_traffic} shows the average rate of the proposed method and the beam alignment using position only in high and low traffic conditions. The datasets used are the same as described in Section \ref{sec:sensitivy_to_traffic_density}. UPA $16\times 16$ and 24 dBm EIRP are assumed. AvgPow is used and the number of beam pairs trained $N_\rmb$ is selected such that $P_\rmpl(\text{3 dB},\mcS)\le 1\%$. The average rates accounting for the beam training overhead and the beam coherence time $T_\rmB$ are computed using \eqref{eq:avg_rate_including_beam_training_overhead} assuming a speed of 20 m/s.  
By using only the position, the beam training can be eliminated, but the method only works when the LOS path is available. This means that it will perform poorly when blockage of the LOS path occurs frequently. We can confirm this effect in Fig.~\ref{fig:avg_rate_comparison_posi_only_diff_traffic}. In the high traffic case, there are many trucks on the street which often block the LOS directions leading to a large gap compared to the proposed method. The gap becomes smaller in the low traffic case because of the less blockage. 
Leveraging the fingerprints database, the proposed method works in all traffic conditions. The benefit of fingerprints increases with the traffic density, or more generally the blockage probability of the LOS path. 

\subsection{Online Data Collection for Type B Fingerprints} \label{sec:online_learning}
While we described in this paper an offline method for database collection, online approaches are also feasible. Here, we provide an example implementation of the online database collection when using Type B fingerprints. This is meant to be an example, and further research is required to design an optimal exploration-exploitation tradeoff. 

We now describe the proposed heuristic for online database collection. We assume the system starts with an initial database of some small size $N_{\mathrm{init}}$ (recall that this is the number of rows in Table \ref{tab:raw_fingerprint_example}), which is obtained from $N_\mathrm{init}$ full exhaustive searches. Because $N_\mathrm{init}$ is small, the initial database is not accurate. The purpose of the initial database is to screen relevant beam pairs that should be learned. The initial database is built in this way. For the first exhaustive search, we keep the top-200 beam pairs (ranked by their received powers). For subsequent exhaustive searches, we examine the top-200 beam pairs. If the beam pairs have already been seen, we update the average received powers of those beam pairs and if they have not yet been seen, we add the beam pairs into the database. In this way, we obtain the initial database that looks something like Table \ref{tab:fingerprint_example}. For the online operation, we assume each vehicle will measure $N_\rmb$ beam pairs, which consist of $N_\mathrm{explore}$ for helping collect data for updating the database and $N_\mathrm{exploit}$ for the vehicle's own beam alignment purpose. As the measurement data accumulate, the exploration can be decreased. We assume a linearly decreasing function for this purpose. Let $t$ be the time index indicating the number of times $N_\rmb$ beam measurements have been conducted, $r_\rminit$ and $\epsilon$ be the exploration ratio in the initial stage and when the database becomes accurate enough, $\xi$ be the rate of decrease in the exploration, and $\lceil\cdot \rceil$ denote the ceiling function, we define the function as
\begin{align}
N_\mathrm{explore}[t] = \begin{cases}
\lceil (r_\mathrm{init}-\xi t)\times N_\rmb  \rceil & \text{if } r_\mathrm{init}-\xi t \ge \epsilon \\
\lceil \epsilon\times N_\rmb  \rceil & \text{if } r_\mathrm{init}-\xi t < \epsilon
\end{cases}.
\end{align}
Note that $N_\mathrm{explore}[t]+N_\mathrm{exploit}[t]=N_\rmb$ for all $t$. The $N_\mathrm{exploit}[t]$ beam pairs are chosen using the AvgPow beam selection method and the database available at time $t$. The $N_\mathrm{explore}[t]$ beams are selected randomly among the beam pairs in the current database (excluding the $N_\mathrm{exploit}[t]$ pairs already selected). This random exploration among the different choices is inspired by the $\epsilon$-greedy approach used in reinforcement learning problems \cite{Sutton1998}. 

\begin{figure}
\centering
\includegraphics[width=\columnwidth]{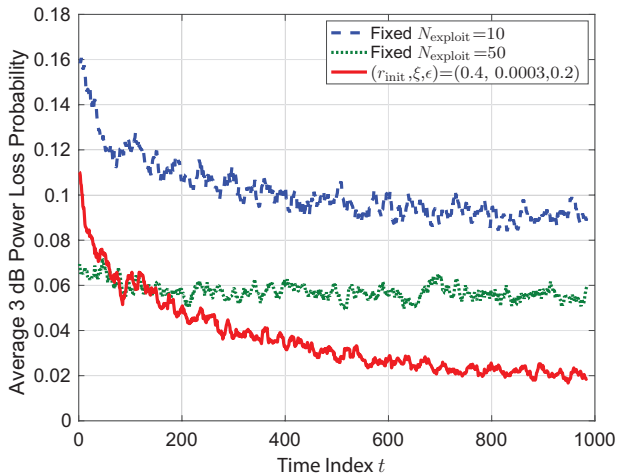}
\caption{Average 3 dB power loss probability when collecting the fingerprint database online. When exploring too much (Fixed $N_\mathrm{exploit}=10$), the system cannot take advantage of the available database. When exploiting too much (Fixed $N_\mathrm{exploit}=50$), the accuracy of the database cannot be improved. It is important to balance the exploration-exploitation tradeoff for a better performance in the long run as shown in the case when setting the tradeoff parameters as  $(r_\mathrm{init},\xi,\epsilon)=(0.4,0.0003,0.2)$.}
\label{fig:online_learning_explore_exploit_tradeoff}
\end{figure}

Finally, we show a numerical example to show the effect of the exploration-exploitation tradeoff. For the evaluation, we use the 1000 samples as in Section \ref{sec:effect_bin_size}. To eliminate the dependence on the order of the data, we randomly permute the 1000 samples and run the online beam alignment 500 times. The metric for the evaluation is the 3 dB power loss probability averaged over the 500 runs. To better show the trend we perform smoothing average (window size of 10 indices) on the time index $t$ on the raw average 3 dB power loss probability. 
Fig. \ref{fig:online_learning_explore_exploit_tradeoff} shows an example of exploration-exploitation tradeoff. In this example, the total number of beam pairs trained per vehicle $N_\rmb=50$ and the initial database size $N_{\mathrm{init}}=5$ are used. We show the case of too-much-exploration (fixed $N_\mathrm{exploit}=10$), too-much-exploitation (fixed $N_\mathrm{exploit}=50$), and a balanced exploration-exploitation ($r_\mathrm{init}=0.4,\xi=0.0003$, and $\epsilon=0.2$). When exploring too much, the system cannot take advantage of the database, and when exploiting too much, the accuracy of the database cannot be improved. Balancing the two provides a better performance in the long run. We note that this example is meant to show the feasibility of an online implementation. Optimal exploration-exploitation tradeoff will depend on how to balance the burden of exploration among the users at different stages of the system.

\section{Conclusion} \label{sec:conclusion} 
We proposed an efficient beam alignment method for mmWave V2I communications leveraging position information and multipath fingerprints. Two types of fingerprints tailored with two beam pair selection methods were proposed. The results show that when Type A fingerprints are available, which can capture correlation between beam pairs, minimizing the misalignment probability is the best method for beam selection. 
If Type A is not possible, then our proposed heuristic beam selection by ranking the average received power should be used with Type B fingerprints, which only store the average received power.
Regarding the overhead of the beam training, our approach requires training less than 30 beam pairs when 16$\times$16 arrays are used and the overhead increases roughly linearly with the number of antenna elements. The overhead calculation under the mobility context using the beam coherence time shows that while the proposed approach consumes less than a few percent of the beam coherence time for training, the IEEE 802.11ad beam training duration can exceed the beam coherence time for large arrays such as 32$\times$32. It is also shown that when adaptation to traffic conditions is not possible, fingerprints should be collected during dense traffic conditions so that most possible paths can be captured in the database. This work demonstrates that side information can be exploited to improve the efficiency of mmWave communications, which is not only desirable but also necessary in vehicular settings.


%


%
%
%

\balance

\ifCLASSOPTIONcaptionsoff
  \newpage
\fi

\bibliographystyle{IEEEtran}
\bibliography{ref,IEEEabrv}

%

%
%
%




\end{document}